\documentclass[pra, twocolumn,superscriptaddress,nofootinbib]{revtex4-2}

\usepackage[english]{babel}
\usepackage[utf8]{inputenc}
\usepackage[T1]{fontenc}
\usepackage{float}
\usepackage{placeins}  % in preamble
\usepackage{booktabs}
\usepackage{multirow}
\usepackage{siunitx} % For professional alignment of numbers
\usepackage{amsmath,amssymb}
\usepackage{graphicx}
\usepackage{tikz}   
\usetikzlibrary{quantikz2,positioning}
\usepackage{longtable}
\usepackage{booktabs}
\usepackage{adjustbox}
\usepackage{subfigure}
\usepackage{tabularx}
\usepackage[colorlinks=true, allcolors=blue]{hyperref}

\begin{document}
% \linenumbers

% \title{Effective operator sorting-based quantum selected configuration interaction: Application to permanent electric dipole moment of molecules}

\title{Resource-efficient quantum-selected configuration interaction for molecular properties}

\author{Suprava Sahoo}
\email{supravasahoo.phys@gmail.com}
\affiliation{Centre for Quantum Engineering, Research and Education, TCG CREST, Kolkata 700091, India}
\affiliation{Academy of Scientific and Innovative Research (AcSIR), Ghaziabad 201002, India}

\author{Abdul Kalam}
% \email{ak6994231@gmail.com}
\affiliation{Centre for Quantum Engineering, Research and Education, TCG CREST, Kolkata 700091, India}
\affiliation{Academy of Scientific and Innovative Research (AcSIR), Ghaziabad 201002, India}

\author{Kenji Sugisaki}
% \email{ak6994231@gmail.com}
\affiliation{Deloitte Tohmatsu LLC, 3-2-3 Marunouchi, Chiyoda-ku, Tokyo 100-8363, Japan}
\affiliation{Centre for Quantum Engineering, Research and Education, TCG CREST, Kolkata 700091, India}

\author{V. S. Prasannaa}
\affiliation{Centre for Quantum Engineering, Research and Education, TCG CREST, Kolkata 700091, India}
\affiliation{Academy of Scientific and Innovative Research (AcSIR), Ghaziabad 201002, India}
\author{B. P. Das}
\affiliation{Centre for Quantum Engineering, Research and Education, TCG CREST, Kolkata 700091, India}
\affiliation{Department of Physics, Institute of Science Tokyo, 2-12-1 Ookayama, Meguro-ku City, Tokyo 152-8550, Japan}

\begin{abstract} 
% The quantum-selected configuration interaction identifies important determinantal basis sets and construct hamiltonian basis and diagonalize it. 

% through real-time evolution of a reference wavefunction; however, simulating the full electronic Hamiltonian on noisy quantum devices results in rapidly growing circuit complexity, limiting the scalability of the method.
The quantum-selected configuration interaction identifies important determinantal basis functions through real-time evolution of a reference wavefunction and diagonalizing the Hamiltonian matrix in the resulting selected subspace. However, implementing the full electronic Hamiltonian on noisy quantum devices leads to rapidly increasing circuit complexity, limiting its scalability.
To address this issue, we identify the dominant fermionic excitation operators and perform reference-state fidelity loss analysis to construct a compact Hamiltonian, reducing computational overhead while retaining high precision.
Applied to Group IIIA monofluorides (BF, AlF, GaF, InF, and TlF), the proposed framework achieves a near-quadratic improvement in Hamiltonian-term scaling, enabling resource-efficient simulations.
We employ this framework to compute the relativistic ground-state energies and permanent electric dipole moments (PDMs) of the systems under consideration. After validating the framework via simulations, we demonstrate hardware execution for AlF and TlF on the IBM Marrakesh processor using active spaces of up to 20 qubits. For a 20-qubit TlF system, the reduced Hamiltonian yields a reduction of higher than $ 98\%$ in both circuit depth and two-qubit gate counts, with the resulting PDMs from quantum hardware matching complete active space configuration interaction values within \(99.99\%\). These results demonstrate the scalability of this approach on noisy intermediate-scale quantum devices.

\end{abstract} 

\maketitle

%\tableofcontents

\section{Introduction}

Simulations of quantum many-body systems, particularly atoms and molecules, are among the most promising applications of quantum computing~\cite{cao2019quantum,mcardle2020quantum}. Although the full configuration interaction (FCI) method provides exact solutions within a chosen single-particle basis, its exponential scaling with Hilbert-space size makes it classically intractable~\cite{shavitt1977method}. Quantum Phase Estimation offers a polynomial-scaling alternative by extracting Hamiltonian eigenvalues through unitary time evolution, but its large circuit depth currently limits practical implementations to relatively small systems~\cite{aspuru2005simulated,santagati2018witnessing,blunt2023statistical,yamamoto2024demonstrating,mohammadbagherpoor2019improved}. Hybrid quantum-classical algorithms, such as the variational quantum eigensolver (VQE), have therefore emerged as promising approaches for noisy intermediate-scale quantum (NISQ) devices~\cite{peruzzo2014variational}. However, VQE typically requires extensive measurements to achieve high accuracy and is susceptible to optimization challenges, including barren plateaus~\cite{tilly2022variational,gonthier2022measurements,mcclean2018barren}.

Recently, quantum-selected configuration interaction (QSCI)~\cite{kanno2026quantum} has emerged as a promising hybrid quantum-classical approach for molecular physics calculations on current quantum hardware. In QSCI, a quantum state that approximates the ground state is prepared and sampled to identify the most important determinantal basis functions (electronic configurations), which are then used to construct a reduced Hamiltonian that is diagonalized classically. Various approaches have been proposed for preparing the approximate quantum state, including adiabatic state preparation, variational methods, and quantum dynamics~\cite{nakagawa2024adapt,farhi2000quantum,robledo2025chemistry}. Lately, several QSCI variants have employed real-time evolution of a suitable reference state, such as the Hartree-Fock (HF) state, under the target Hamiltonian to generate the input state. These include Hamiltonian simulation-based QSCI (HSB-QSCI)~\cite{sugisaki2025hamiltonian}, time-evolved QSCI (TE-QSCI)~\cite{mikkelsen2025quantum}, and sample-based Krylov quantum diagonalization~\cite{yu2025quantum}. Among these approaches, HSB-QSCI constructs the selected determinant subspace from the sampled configurations and subsequently diagonalizes the projected Hamiltonian classically, thus reducing quantum noise sensitivity while avoiding the need for highly accurate time evolution, controlled operations, or ancillary qubits~\cite{mikkelsen2025quantum}.

The quantum resource requirements of HSB-QSCI increase polynomially with the system size.
For a system of $N$ qubits, a single Trotter step generates $\mathcal{O}(N^{4})$ Hamiltonian Pauli terms, resulting in $\mathcal{O}(N^{4})$ single-qubit rotation gates and $\mathcal{O}(N^{5})$ two-qubit (CX) gates under the Jordan-Wigner transformation~\cite{mikkelsen2025quantum}. The resulting circuit depth poses a major challenge for NISQ devices. Moreover, the measurement process for the full time-evolved circuit is inherently sequential and cannot be parallelized. After measurement, the sampled configurations define a subspace Hamiltonian of dimension $D\times D$, and the subsequent classical diagonalization scales as $\mathcal{O}(D^{3})$.

QSCI has been successfully applied to ground-state energies~\cite{kanno2026quantum,sugisaki2025hamiltonian,nakagawa2024adapt,mikkelsen2025quantum,robledo2025chemistry,chen2026neural}, quasiparticle band structures~\cite{ohgoe2025quantum}, excited-state energies of biomolecules~\cite{yamamoto2026quantum}, and binding energies of weakly bound molecular systems~\cite{kaliakin2025accurate,sugisaki2025size}. However, its application to molecular properties beyond energies, such as permanent electric dipole moments (PDMs), remains limited, and its extension to relativistic electronic-structure problems has received little attention. Since relativistic effects play a crucial role in determining the chemical and physical properties of many molecules, incorporating them into quantum algorithms is essential but remains challenging~\cite{veis2012relativistic,stetina2022simulating,chawla2025relativistic,kumar2024computation,singh2024experimental,sugisaki2023bayesian}. In this work, we address both challenges by extending the QSCI framework to relativistic calculations of molecular properties beyond ground-state energies.
PDMs govern long-range dipole-dipole interactions in cold and ultracold molecular systems, enabling applications in quantum technologies and the exploration of novel quantum phases~\cite{mishra2009supersolid,trefzger2011ultracold,demille2002quantum}. They are also useful for probing fundamental physics~\cite{vutha2010search}. In particular, Group IIIA monofluorides have gained importance in quantum science because of their stability, relevance to laser cooling and precision measurements, pronounced relativistic effects in heavier systems, and their importance in fundamental physics~\cite{hoeft1970production,padilla2025magneto,hoeft1970microwave,al2016theoretical,truppe2019spectroscopic,hunter2012prospects,abe2020accurate,grasdijk2021centrex,gao2015laser,yang2017feasibility}.

To address these limitations, we introduce an effective Pauli operator sorting-based QSCI (eos-QSCI) framework that reduces Hamiltonian complexity before quantum simulation by identifying and retaining only the dominant Pauli operator terms. This constructs a compact, effective Hamiltonian, thereby substantially reducing the circuit complexities. It also reduces the determinant space sampled in HSB-QSCI while preserving its accuracy. Unlike approaches that primarily compress quantum circuits through specialized tensor-network optimization techniques~\cite{pan2022simulation,rogerson2024quantum}, eos-QSCI reduces the computational cost at the Hamiltonian level, making the subsequent time evolution inherently more efficient.
Relativistic electronic-structure calculations require essentially more Hamiltonian terms than non-relativistic calculations, increasing quantum resource demands~\cite{chawla2025relativistic}. We apply eos-QSCI to Group IIIA monofluorides and demonstrate up to 20-qubit quantum hardware calculations for AlF and TlF on IBM's Marrakesh processor. We benchmark the method using ground-state energies and PDMs, while evaluating quantum and classical resource savings against full-Hamiltonian HSB-QSCI.

\section{Theory} 

\subsection{Relativistic Hamiltonian and PDM} 

To account for relativistic effects in molecular systems, we employ the Dirac–Coulomb Hamiltonian in the Born–Oppenheimer approximation, which is given by (see Chapter 6 of~\cite{grant2007relativistic}) 
\begin{equation}
H_{DC} = \sum_{i=1}^{N_e} \left[ c\,\boldsymbol{\alpha}_i \cdot \mathbf{p}_i + \beta_i c^2 + V_{\mathrm{nuc}}(\mathbf{r}_i) \right]
+ \frac{1}{2} \sum_{i \ne j}^{N_e} \frac{1}{|\mathbf{r}_i - \mathbf{r}_j|},
\end{equation}
where $c$ is the speed of light, $\mathbf{p}_i$ is the momentum of the $i^{\text{th}}$ electron, $N_e$ is the number of electrons, and $\mathbf{r}_i$ is the coordinate vector of the $i^{\text{th}}$ electron. The quantities $\boldsymbol{\alpha}_i$ and $\beta_i$ are the Dirac matrices.

$V_{\mathrm{nuc}}(\mathbf{r}_i)$ is the electron--nucleus potential of the $i^{\text{th}}$ electron. We use atomic units (a.u.) throughout. 

In heteronuclear diatomic molecules, the unequal distribution of electron density between the two nuclei leads to a finite PDM. For an electronic state $|\Psi\rangle$ it is evaluated as, $\text{PDM} = \frac{\langle \Psi | D | \Psi \rangle}{\langle \Psi | \Psi \rangle}$. Considering only the electronic contribution part, this operator can be expressed in the qubit representation as $D = \sum_n d_n P_n,$
where $d_n$ are coefficients determined by the PDM one-electron integrals, and $P_n$ denote the Pauli strings obtained from the transformation of the dipole moment operator from its second-quantized form to the qubit operator form. Since the electronic PDM calculations are performed within an active space, the total PDM is obtained by adding the frozen-core electronic and nuclear contributions to the computed value.
%%%%%%%%%%%%%%%%%%%%%%%%%%%%%%%%%%%%%%%%%%%%%%%%%%%%%%%%%%%%%%%%%%%%%%%%%%%%%%%%%%%%%%%%%%%%%%%%%%%

\subsection{Configuration interaction method}

Next, we briefly review the configuration interaction (CI) framework underlying the QSCI method. Within this framework, the many-electron wavefunction is expanded in a basis of orthonormal Slater determinants (or configuration state functions)~\cite{szabo2012modern}.
Using the Hartree--Fock (HF) determinant $\ket{\Phi_0}$ as the reference, the CI wavefunction may be written in second-quantized form as
\begin{equation}
\begin{aligned}
    \ket{\Psi} &= c_0 \ket{\Phi_0}
    + \sum_{i,a} c_i^a \ket{\Phi_i^a}
    + \sum_{i<j,\,a<b} c_{ij}^{ab} \ket{\Phi_{ij}^{ab}}
    + \cdots \\
    &= \left(
    c_0 
    + \sum_{i,a} c_i^a a_a^\dagger a_i
    + \sum_{i<j,\,a<b} c_{ij}^{ab} a_a^\dagger a_b^\dagger a_j a_i
    + \cdots
    \right)\ket{\Phi_0},
\end{aligned}
\end{equation}
where $a_p^\dagger$ and $a_q$ are fermionic creation and annihilation operators, respectively. Throughout this work, $i,j,\cdots$ denote occupied orbitals, $a,b,\cdots$ denote virtual orbitals, and $p,q,r,s$ denote general orbitals.
The coefficients ${c_0, c_i^a,c_{ij}^{ab},\cdots }$ are obtained by solving the CI secular equation, $H\mathbf{c}=E\mathbf{c}$. Inclusion of all possible excitations yields the FCI limit.

The exponential growth of the FCI determinant space makes exact calculations computationally prohibitive. Selected configuration interaction addresses this challenge by retaining only the dominant determinants~\cite{holmes2016heat,kanno2026quantum}. QSCI extends this concept by identifying these determinants through quantum state preparation and measurement.

\subsection{Hamiltonian Simulation-Based QSCI}

In HSB-QSCI, the selected determinants are sampled from the real-time evolution of an approximate wavefunction. This begins from a reference state $|\Phi_0\rangle$ with a large overlap with the target state and the second-quantized electronic Hamiltonian,

\begin{figure*}[t]
    \centering
    \includegraphics[width=0.8\textwidth]{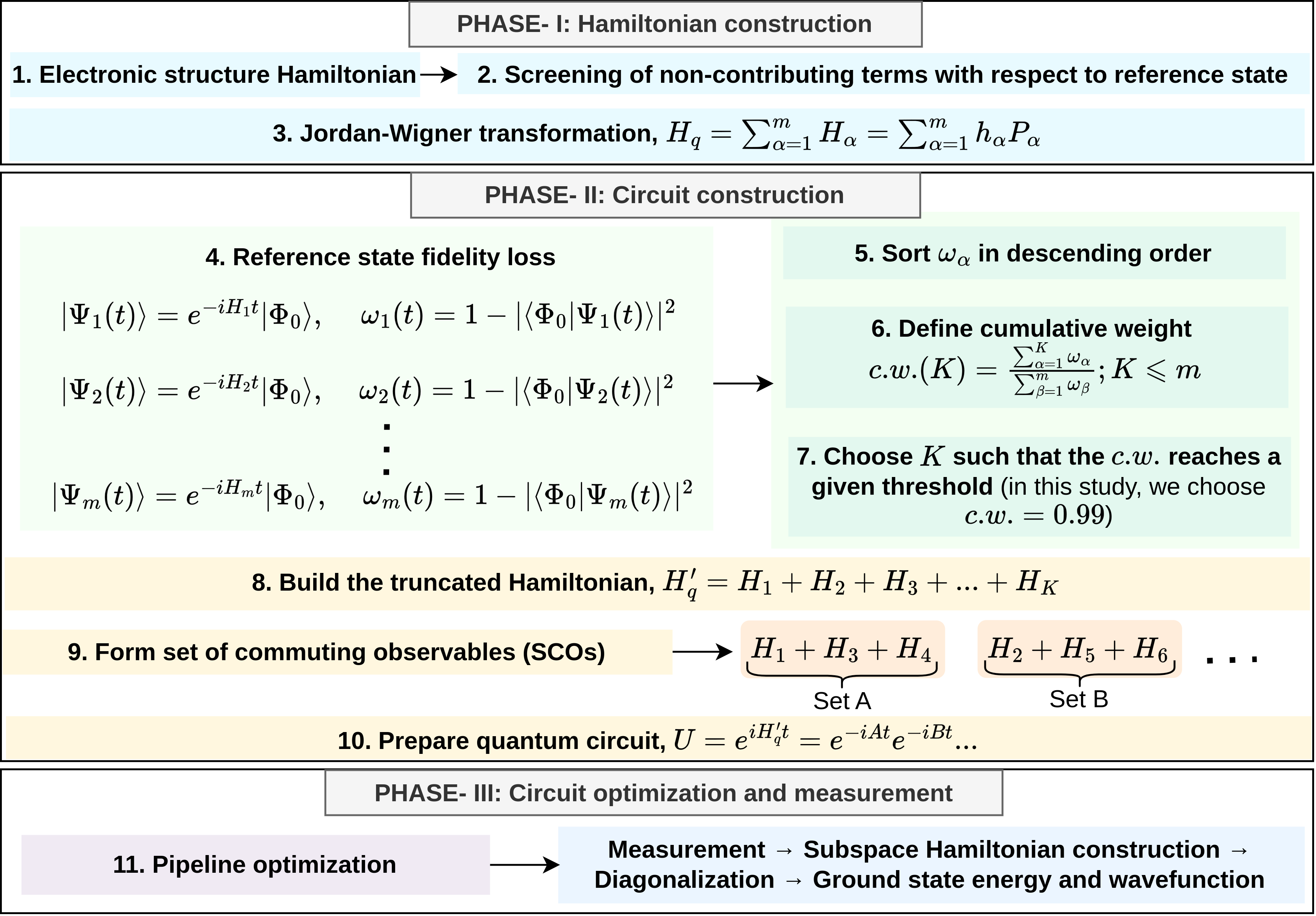}
    \caption{Schematic workflow of the proposed eos-QSCI method. \textbf{PHASE I:} Construction of the truncated Hamiltonian through reference-state-based screening of non-contributing terms followed by the Jordan–Wigner transformation. \textbf{PHASE II:} Time-evolution circuit construction via fidelity loss and cumulative-weight-based Hamiltonian-term selection and grouping into sets of commuting observables (SCOs). \textbf{PHASE III:} Pipeline optimization, quantum hardware execution, and classical post-processing to obtain the ground-state energy and wavefunction.
}
    \label{fig:flowchart}
\end{figure*}

%%%%%%%%%%%%%%%%%%%%%%%%

\begin{equation}
H = \sum_{pq} h_{pq}\, a_p^\dagger a_q
+ \frac{1}{2} \sum_{pqrs} g_{pqrs}\, a_p^\dagger a_q^\dagger a_s a_r,
\label{eq:second_quantized_hamiltonian}
\end{equation}
where $h_{pq}$ and $g_{pqrs}$ are the one- and two-electron integrals respectively. Applying the Jordan--Wigner transformation~\cite{jordan1928paulische} yields the qubit Hamiltonian
\begin{equation}
H_q=\sum_{\alpha=1}^{\tilde{m}}h_\alpha P_\alpha,
\label{qubit_op}
\end{equation}
where $P_\alpha \in\left\{ I,X,Y,Z\right\}^{\bigotimes N}$ are Pauli operators with coefficients $h_\alpha$.
The quantum state is generated through a sequence of real-time evolutions,
\begin{equation}
|\Phi_k\rangle=e^{-iH_q \mathcal{K}t}|\Phi_0\rangle, \qquad \mathcal{K}=1,2,\cdots,k
\label{Eq:time_evol_state}
\end{equation}
implemented using Hamiltonian simulation techniques such as the Trotter--Suzuki decomposition~\cite{trotter1959product,suzuki1976relationship}, where $\mathcal{K}$ is the Krylov expansion order. The total evolution time length $\mathcal{K} t$ governs quantum interference among eigenstates, thereby determining the important Slater determinants sampled in the computational basis. Measurements of $|\Phi_k\rangle$ in the computational basis sample the dominant configurations that define the reduced Hamiltonian, which is diagonalized classically.

Thus, HSB-QSCI identifies the most relevant configurations and constructs an orthonormal selected basis from the sampled configurations. This is closely related to quantum Krylov subspace methods~\cite{cortes2022quantum}, where successive real-time evolution generates a non-orthogonal Krylov basis, leading to the generalized eigenvalue problem $Hc=ESc$, where $S$ is the overlap matrix. In contrast, HSB-QSCI directly samples orthonormal Slater determinants, reducing the problem to the standard eigenvalue equation $Hc=Ec$ while avoiding overlap-matrix conditioning issues.

The first-order Trotter-Suzuki product formula is
\begin{equation}
e^{-i\sum_\alpha H_\alpha t} \approx
\lim_{n\rightarrow\infty}
\left(
\prod_{\alpha=1}^{m}
e^{-iH_\alpha t/n}
\right)^n,
\label{eq:trotter}
\end{equation}
with Trotter error scaling as, $\varepsilon=\mathcal{O}\left(\frac{t^2}{n}\right)$~\cite{childs2021theory}.
Increasing Trotter step $n$ reduces the simulation error, but proportionally increases circuit depth, since each step requires implementing all non-commuting Hamiltonian terms. Randomized Hamiltonian simulation methods, such as qDRIFT~\cite{campbell2018random}, reduce circuit cost by stochastically sampling Hamiltonian terms. In contrast, the proposed eos-QSCI framework reduces computational resources by constructing a compact effective Hamiltonian through operator screening.

\subsection{eos-QSCI method}

Fig.~\ref{fig:flowchart} summarizes the eos-QSCI framework, which consists of three phases. PHASE-I and PHASE-II perform Hamiltonian operator screening and selection, while PHASE-III compiles the resulting circuits for quantum execution and carries out the HSB-QSCI calculations.

In PHASE-I, the second-quantized Hamiltonian [Eq.~\eqref{eq:second_quantized_hamiltonian}] is screened by examining the action of each fermionic operator string $\hat{\Omega}$ on the reference determinant $\ket{\Phi_0}$. Only operators satisfying
$\hat{\Omega}\ket{\Phi_0}\neq0$ are retained, thereby imposing occupation-based selection rules. For one-body operators, $\hat{\Omega}_{pq}=a_p^\dagger a_q$, a nonzero contribution requires the annihilation index $q\in{i,j,\ldots}$, yielding the single excitations, $a_a^\dagger a_i\ket{\Phi_0}=\ket{\Phi_i^a}.$
Similarly, for two-body operators, $\hat{\Omega}_{pqrs}=a_p^\dagger a_q^\dagger a_s a_r$, both annihilation indices, $(r,s)\in{i,j,\ldots}$, producing the double excitations, $a_a^\dagger a_b^\dagger a_j a_i\ket{\Phi_0}=
\ket{\Phi_{ij}^{ab}},$
whereas all operators containing annihilation on virtual orbitals vanish. The Hermitian conjugate of every retained excitation operator is also included. This occupation-based screening provides an efficient first-stage reduction of the Hamiltonian before the subsequent importance-based truncation (see Appendix A of the Supplemental Material).
The screened fermionic Hamiltonian is then mapped to qubit Hamiltonian through the Jordan-Wigner transformation, yielding a Pauli expansion of the form in Eq.~\eqref{qubit_op} containing $m$ terms, where $m<\tilde{m}$.

In PHASE-II, the retained Pauli operators are ranked according to their ability to excite the reference state. We quantify the importance of each Pauli operator using the reference-state fidelity loss, where the reference state $|\Phi_0\rangle$ is the Dirac--Hartree--Fock (DHF) ground state (Chapter 6 of~\cite{grant2007relativistic}). For each Pauli operator $P_\alpha$, we consider the short-time evolution generated solely by that operator,
\begin{equation}
|\Psi_\alpha(t)\rangle=
e^{-ih_\alpha P_\alpha t}
|\Phi_0\rangle.
\end{equation}
Expanding to second order in $t$ gives
\begin{equation}
|\Psi_\alpha(t)\rangle=
|\Phi_0\rangle
-ih_\alpha P_\alpha t|\Phi_0\rangle
-\frac{(h_\alpha P_\alpha t)^2}{2}|\Phi_0\rangle
+\cdots,
\end{equation}
from which the DHF fidelity loss is
\begin{equation}
\begin{aligned}
\omega_{\alpha}(t)
&=
1-|\langle \Phi_0|\Psi_\alpha(t)\rangle|^2 \\
&\approx
h_\alpha^2 t^2
\left[
1-\big|\langle \Phi_0|P_\alpha|\Phi_0\rangle\big|^2
\right]
\end{aligned}
\end{equation}
Thus, $\omega_\alpha(t)$ is proportional to the variance of $P_\alpha$ with respect to the DHF reference state and serves as an effective measure of its importance.
In the present work, $\omega_\alpha(t)$ is evaluated classically due to current quantum hardware limitations, although it should be ideally executed on a quantum device. Since each circuit contains at most $2(N-1)$ two-qubit gates, a high-fidelity implementation is expected. Moreover, the circuits are independent and can be executed in parallel across distributed quantum resources.

The Pauli operators are sorted in descending order of $\omega_\alpha$, and a cumulative normalized weight is defined as
\begin{equation}
c.w.(K)=
\frac{\sum_{\alpha=1}^{K}\omega_\alpha}
{\sum_{\beta=1}^{m}\omega_\beta},
\qquad
K\le m,
\label{cw(K)}
\end{equation}
where $K$ is the number of retained Hamiltonian terms. The truncation parameter $K$ is chosen as the smallest value satisfying $c.w.(K)\gtrsim0.99$. Operators with larger $\omega_\alpha$ induce greater deviations from the DHF reference state; therefore they are expected to excite the DHF state and contribute more significantly to the time-evolved wavefunction.

The present framework ranks Hamiltonian terms based on their action on the reference state, $U^\mathcal{K}|\Phi_0\rangle$ for $\mathcal{K}=1$. For $\mathcal{K}>1$, the evolved state becomes multi-configurational, allowing operators that initially have negligible contributions to become important through higher-order excitations. The systematic assessment of this limitation is left for future investigation.

The selected operators define the truncated Hamiltonian,
\begin{equation}
H_q'=\sum_{\alpha=1}^{K}h_\alpha P_\alpha,
\end{equation}
which is partitioned into sets of commuting observables (SCOs)~\cite{brady2024iterative,verteletskyi2020measurement} to construct the time-evolution circuit group-wise.

In PHASE-III, the resulting circuit is further optimized using pipeline optimization routines: Qiskit (L3)~\cite{javadi2024quantum} and Pytket (multiple-sequence pass)~\cite{sivarajah2021t}, reducing both circuit depth and CX gate count. The optimized circuits are then executed within the HSB-QSCI framework to obtain the ground-state energy and wavefunction, from which PDMs are evaluated.

\section{Results and discussion}

\begin{figure}[t]
    \centering
    \includegraphics[width=\columnwidth]{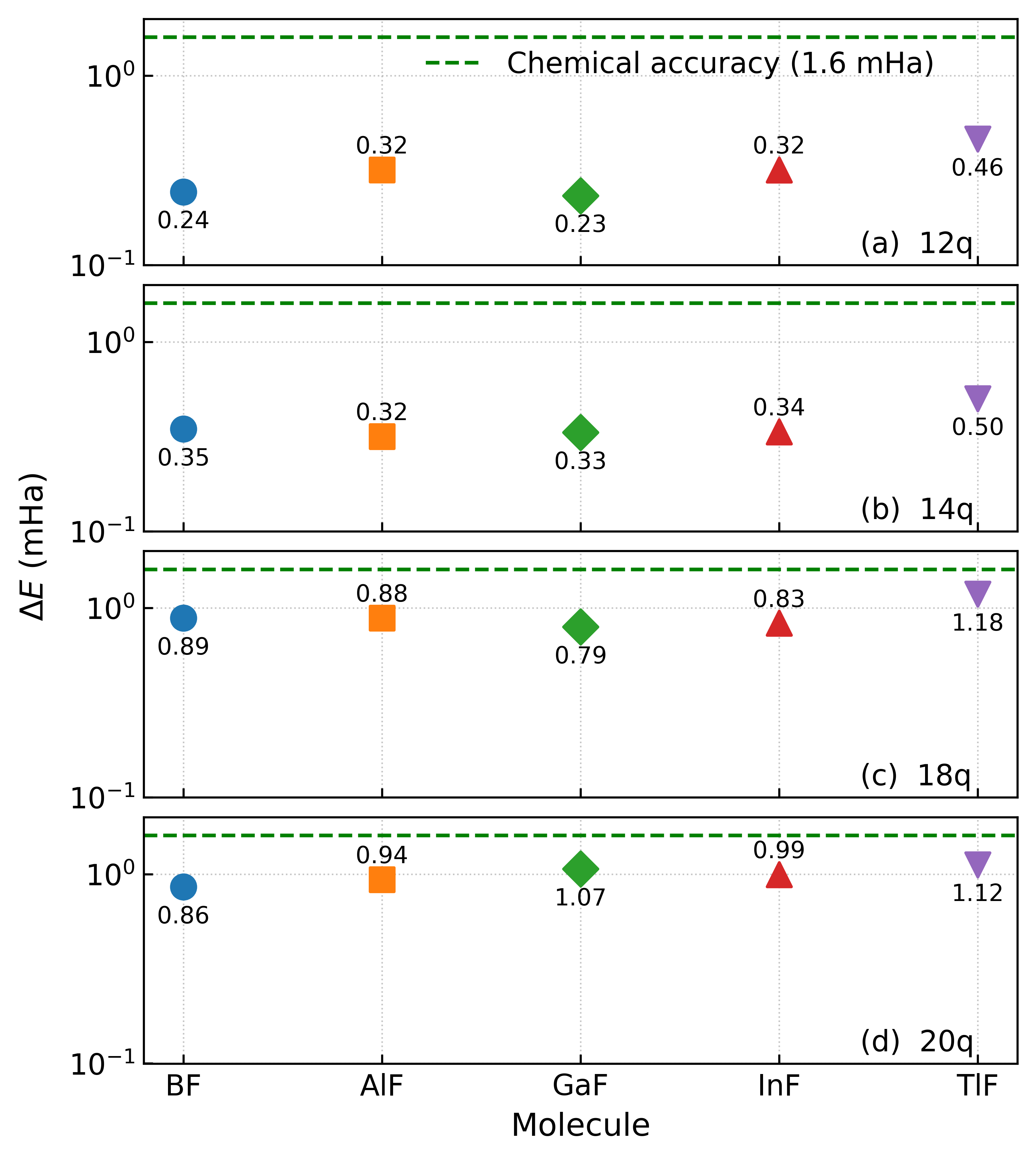}
\caption{Absolute energy deviations, $\Delta E = |E_{\mathrm{eos\text{-}QSCI}}-E_{\mathrm{CASCI}}|$, for Group IIIA monofluorides obtained with eos-QSCI for active spaces corresponding to (a) 12, (b) 14, (c) 18, and (d) 20 qubits. The vertical axis is shown on a logarithmic scale. The dashed horizontal line indicates the chemical accuracy threshold (1.6 mHa), and the numerical values of $\Delta E$ (mHa) are labeled at each data point.}
    \label{fig:energy-difference}
\end{figure}

\begin{table*}[t]
\centering
\small
\setlength{\tabcolsep}{4pt}
\renewcommand{\arraystretch}{1.15}

\caption{
Reduction in Hamiltonian terms, circuit depth, and CX gate count achieved by eos-QSCI for AlF and TlF at $c.w.(K)=0.99$ across different system sizes. Percentage reductions are computed as, $\%\mathrm{Red.}=\frac{\mathrm{Full}-\mathrm{eos}}{\mathrm{Full}}\times100,$ where \textbf{Full} and \textbf{eos}  denote the resource counts obtained with the full Hamiltonian and eos-QSCI, respectively.
}

\label{tab:alf_tlf_resources}

\begin{ruledtabular}
\begin{tabular}{c c|ccc|ccc|ccc}
\multirow{2}{*}{Molecule}
& \multirow{2}{*}{System size}
& \multicolumn{3}{c|}{Hamiltonian term}
& \multicolumn{3}{c|}{Circuit depth}
& \multicolumn{3}{c}{CX gate count} \\

& 
& Full
& eos
& \% Red.
& Full
& eos
& \% Red.
& Full
& eos
& \% Red. \\

\hline

\multirow{4}{*}{AlF}

& 12
& 1 563
& 79
& 94.94
& 20 968
& 669
& 96.80
& 14 276
& 603
& 95.77 \\

& 14
& 2 998
& 129
& 95.69
& 44 401
& 1 468
& 96.69
& 31 558
& 1 265
& 95.99 \\

& 18
& 8 160
& 262
& 96.78
& 141 513
& 3 495
& 97.53
& 106 706
& 3 291
& 96.91 \\

& 20
& 10 931
& 312
& 97.14
& 199 307
& 4 529
& 97.72
& 152 284
& 4 245
& 97.21 \\

\hline

\multirow{4}{*}{TlF}

& 12
& 1 757
& 115
& 93.45
& 24 021
& 948
& 96.05
& 16 472
& 809
& 95.08 \\

& 14
& 3 424
& 166
& 95.15
& 51 662
& 1 570
& 96.96
& 36 818
& 1 376
& 96.26 \\

& 18
& 9 302
& 265
& 97.15
& 163 192
& 3 486
& 97.86
& 123 166
& 3 241
& 97.36 \\

& 20
& 13 583
& 271
& 98.00
& 252 679
& 3 600 
& 98.57
& 194 436
& 3 333
& 98.28 \\

\end{tabular}
\end{ruledtabular}
\end{table*}

All calculations employ the dyall.v4z basis set~\cite{dyall2006relativistic} and optimized equilibrium geometries from Refs.~\cite{murad1966dissociation,jaszunsnki2014spin}: BF (1.2625 Å), AlF (1.6543 Å), GaF (1.7743 Å), InF (1.9854 Å), and TlF (2.0844 Å). Relativistic four-component one- and two-electron integrals are generated using DIRAC25~\cite{saue2020dirac} and mapped to qubit Hamiltonians in Qiskit 1.4.1~\cite{javadi2024quantum} through the DIRAC–OpenFermion interface~\cite{saue2020dirac}. The one-electron integrals required for PDM calculations are also obtained from DIRAC25.

Active spaces of $(4e,6o)$, $(4e,7o)$, $(4e,9o)$, and $(4e,10o)$ are considered, corresponding to 12-, 14-, 18-, and 20-qubit Hamiltonians, where $e$ and $o$ denote the numbers of active electrons and spatial orbitals, respectively. The computed ground-state energies and PDMs are benchmarked against complete active space configuration interaction (CASCI) results, i.e., FCI within the chosen active space. Time evolution is implemented using a first-order Trotter decomposition with a single Trotter step ($n=1$), time $t=1$, and $\mathcal{K}=1$. The resulting projected Hamiltonian is diagonalized classically using the Davidson algorithm~\cite{davidson197514,crouzeix1994davidson}.

% \subsection{Simulation results}

We now focus on the quantum simulation results for the molecular systems under consideration. We first compare the quantum resource requirements of the full-Hamiltonian approach (the complete second-quantized Hamiltonian is directly mapped to a qubit Hamiltonian for time evolution) with eos-QSCI, focusing on the number of Hamiltonian terms, circuit depth, and elementary gate counts required to implement a single time-evolution operator, $e^{-iHt}$.

Table~\ref{tab:alf_tlf_resources} summarizes the quantum resources for AlF and TlF at $c.w.(K)=0.99$. Compared with the full-Hamiltonian approach, the eos-QSCI framework considerably decreases the number of Hamiltonian terms, circuit depth, and CX gate counts. For the 20-qubit AlF system, the number of Hamiltonian terms is reduced by $97.14\%$, leading to a $97.21\%$ reduction in CX gate counts. A similar trend is observed for TlF, where the Hamiltonian terms and CX gate counts are reduced by up to $98.00\%$ and $98.28\%$, respectively. These results demonstrate that the proposed strategy maintains its efficiency even for relatively large 20-qubit relativistic Hamiltonians, significantly lowering the quantum resources required for heavy-element systems.

A higher $c.w.$ threshold retains more Hamiltonian terms, generally increasing the quantum circuit resources. As discussed in Appendix B of the Supplemental Material, the number of retained terms is sensitive to the choice of the $c.w.$ threshold and should therefore be selected appropriately for the system under consideration. For completeness, the corresponding resource statistics, including Hamiltonian terms, optimized circuit complexities, and their percentage reductions for the remaining molecules, are provided in Appendix D of the Supplemental Material.

\begin{table}[!b]
\centering
\small
\setlength{\tabcolsep}{4pt}
\renewcommand{\arraystretch}{1.1}
\caption{Number of determinants (D) sampled from the Statevector simulator for the TlF molecule at different system sizes. The HSB-QSCI with the full Hamiltonian and the eos-QSCI methods substantially reduce the determinant space compared with CASCI.}
\label{tab:determinants}
\begin{tabular}{c c c c c}
\hline\hline
System size & Shots & \multicolumn{3}{c}{Number of determinants (D)} \\
\cline{3-5}
 &  & CASCI & HSB-QSCI & eos-QSCI \\
\hline
12 & $10^5$ & 495 & 22 &  10 \\
14 & $10^5$ & 1001 & 33 & 16 \\
18 & $10^6$ & 3060 & 121 & 32 \\
20 & $2\times10^6$ & 4845 & 181 & 43 \\
\hline\hline
\end{tabular}
\end{table}

Besides reducing quantum circuit complexity, the Hamiltonian truncation also affects the size of the determinant space generated during HSB-QSCI, which governs the cost of the classical diagonalization. Table~\ref{tab:determinants} lists the sampled determinants for TlF obtained using the Statevector simulator. As the system size increases from 12 to 20 qubits, the CASCI Hilbert space grows up to 4845 determinants. In contrast, eos-QSCI requires only $D'$ determinants ($D'<D$), reducing the classical diagonalization cost to $\mathcal{O}(D'^3)$. For the 20-qubit system, retaining only $\sim0.88\%$ of the CASCI determinants is sufficient to recover the ground-state energy within chemical accuracy, requiring the diagonalization of a $43\times43$ Hamiltonian matrix instead of the $181\times181$ matrix used in the full-Hamiltonian HSB-QSCI approach.
To maintain reliable sampling for larger Hilbert spaces, the number of measurement shots is increased from $10^5$ (12 and 14 qubits) to $10^6$ (18 qubits) and $2\times10^6$ (20 qubits). These results demonstrate that eos-QSCI concentrates sampling on the dominant determinants, providing reduction in both the effective Hilbert-space dimension and the measurement overhead.

\begin{table}
\setlength{\tabcolsep}{4pt}
\centering
\caption{Recovered correlation energies and PDMs of AlF and TlF computed with eos-QSCI at different system sizes ($c.w.(K)=0.99$). The absolute PDM deviation is defined as $\Delta\mathrm{PDM}=|\mathrm{PDM}_{\rm eos\text{-}QSCI}-\mathrm{PDM}_{\rm CASCI}|$. PDM values are given in atomic units.}
\begin{tabular}{cccccc}
\toprule
\toprule

System size &
Molecule &
$\%E_{\rm corr}$ &
\begin{tabular}{@{}c@{}}$\mathrm{PDM}$\\$_{\rm CASCI}$\end{tabular} &
\begin{tabular}{@{}c@{}}$\mathrm{PDM}$\\$_{\rm eos\text{-}QSCI}$\end{tabular} &
$\Delta$PDM \\

\midrule

\multirow{2}{*}{12}
& AlF & 96.49 & -0.5565 & -0.5480 & 0.0085 \\
& TlF & 90.45 & -1.6774 & -1.6750 & 0.0024 \\

\midrule

\multirow{2}{*}{14}
& AlF & 98.46 & -0.5894 & -0.5745 & 0.0149 \\
& TlF & 92.01 & -1.6831 & -1.6776 & 0.0055 \\

\midrule

\multirow{2}{*}{18}
& AlF & 97.53 & -0.6881 & -0.6796 & 0.0085 \\
& TlF & 90.02 & -1.6964 & -1.6923 & 0.0041 \\

\midrule

\multirow{2}{*}{20}
& AlF & 97.47 & -0.6950 & -0.6976 & 0.0026 \\
& TlF & 90.63 & -1.6966 & -1.6896 & 0.0070 \\

\bottomrule
\bottomrule
\end{tabular}
\label{tab:AlF_TlF_summary}
\end{table}
%%%%%%%%%%%%%%%%%%%%%%%%%%%%%%%%%%%%%%%%%%%%%%

The accuracy of eos-QSCI is assessed by comparing the computed ground-state energies and PDMs with CASCI reference values. Fig.~\ref{fig:energy-difference} shows the energy deviation, $\Delta E = |E_{\mathrm{eos\text{-}QSCI}}-E_{\mathrm{CASCI}}|$, for all molecules and active spaces (12--20 qubits). The dashed green line denotes the chemical precision threshold (1.6 mHa). For all systems, the deviations remain below this threshold, typically ranging from $0.2$ to $1.1$ mHa. The percentage of recovered correlation energy provides a measure of the effectiveness of the eos-selected determinant space in capturing electron correlation. The recovered correlation energies and PDMs for AlF and TlF, together with the corresponding CASCI reference values, are summarized in Table~\ref{tab:AlF_TlF_summary}. For AlF, eos-QSCI recovers $96.49$--$98.46\%$ of the CASCI correlation energy, while the PDM deviations remain within $1.49\times10^{-2}$ a.u. across all active spaces. Similarly, for TlF, the recovered correlation energy ranges from $90.02$ to $92.01\%$, with PDM deviations not exceeding $7.0\times10^{-3}$ a.u., even for the largest active space. The high recovered correlation energies and small PDM deviations indicate that the eos-selected determinant subspace captures the essential electron correlation while accurately reproducing molecular properties with a reduced CI subspace. The corresponding ground-state energies and PDMs for all molecules are provided in Appendix D of the Supplemental Material.
Overall, these results show that eos-QSCI achieves substantial quantum resource savings while maintaining chemical accuracy for both light and heavy relativistic molecular systems, highlighting its potential for NISQ-era quantum simulations.

\begin{figure*}
    \centering
    \includegraphics[width=1\textwidth]{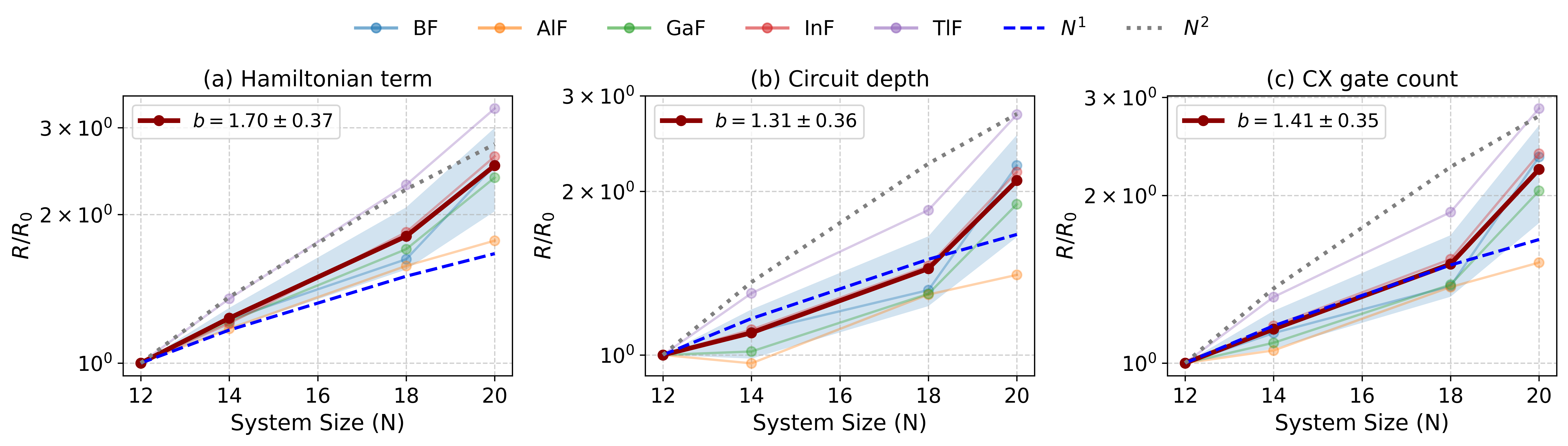}
    \caption{Scaling of the normalized resource reduction ratio, $R/R_0$, with system size $N$ for (a) Hamiltonian terms, (b) circuit depth, and (c) CX gate count. Here, $R(N)=\frac{\mathcal{C}_{\mathrm{Full}}(N)}{\mathcal{C}_{\mathrm{eos}}(N)}$, where $\mathcal{C}$ denotes the corresponding resource metric, normalized to its value at $N_0=12$. Light-colored curves represent individual molecules, while the thick dark-red curve and shaded region denote the mean and one standard deviation, respectively. Power-law fits, $R/R_0\propto N^b$, yield the scaling exponent $b$ (reported in the legend). Dashed blue and dotted gray lines indicate the reference $N^1$ and $N^2$ scaling, respectively.
 }
    \label{fig:scaling}
\end{figure*}

To quantify the reduction in quantum resources, we define the ratio
$R(N)=\frac{\mathcal{C}_{\mathrm{Full}}(N)}{\mathcal{C}_{\mathrm{eos}}(N)}$,
where $\mathcal{C}$ denotes the number of Hamiltonian terms, circuit depth, or CX gate count. This ratio is normalized by its value at the smallest system size ($N_0=12$), and the resulting quantity, $R/R_0$, is plotted as a function of $N$ in Fig.~\ref{fig:scaling}. For all three resource metrics, $R/R_0$ increases with system size, indicating that the resource advantage of eos-QSCI becomes progressively more pronounced as the size of the molecular system grows.

Power-law fits to the normalized ratios yield average scaling exponents of $b=1.70\pm0.37$ for the Hamiltonian terms, $b=1.31\pm0.36$ for the circuit depth, and $b=1.41\pm0.35$ for the CX gate count, with the uncertainty $(\pm)$ indicating one standard deviation of the fitted exponents across the molecular set. The Hamiltonian-term reduction exhibits the greatest improvement, approaching quadratic scaling, while the circuit depth and CX gate counts display approximately linear-to-superlinear scaling. These trends demonstrate that the resource savings provided by eos-QSCI increase systematically with system size. Additional scaling analyses are presented in Appendix C of the Supplemental Material.

Consistent with these observations, the number of retained Pauli terms in eos-QSCI scales as approximately $\mathcal{O}(N^{2.3})$, compared with $\mathcal{O}(N^4)$ for the full-Hamiltonian approach. Since the implementation of each Pauli-string evolution requires approximately $\mathcal{O}(N)$ CX gates, the reduced number of retained terms directly lowers the overall circuit depth and gate count. Consequently, eos-QSCI improves noise resilience and extends the feasible simulation size of relativistic molecular systems on current NISQ hardware.

%%%%%%%%%%%%%%%%%%%%%%%%%%%Hardware Results%%%%%%%%%%%%%%%%%%%%%%%%%%%%%%%%%%%%%
\begin{table*}[t]
\caption{
Quantum hardware results for eos-QSCI calculations of AlF and TlF on the IBM Marrakesh processor. Circuit statistics include the CZ gate count, measurement shots, and the number of sampled determinants before ($\rm D_{\rm noisy}$) and after ($\rm D_{\rm filtered}$) particle-number post-selection and SCCR. $\Delta$ denotes the absolute deviation from the corresponding CASCI value, and $\%E_{\rm corr}$ is the percentage of CASCI correlation energy recovered by eos-QSCI.
}
\label{tab:hardware_summary}
\begin{ruledtabular}
% \footnotesize
\begin{tabular}{ccccccccccc}
Molecule &
Qubits &
CZ &
Shots &
$\text{D}_{\rm noisy}$& $\text{D}_{\rm filtered}$ &
Energy (Ha) &
$\Delta E$ (mHa) &
$\%E_{\rm corr}$ &
PDM (a.u.) &
$\Delta$PDM (a.u.) \\
\hline

\multirow{3}{*}{AlF}
& 12 & 200 & 20000 & 1969 & 135
& $-342.041613$
& 0.034
& 99.62
& $-0.5568$
& 0.0003 \\

& 14 & 1072 & 50000 & 14619 & 386
& $-342.053001$
& 0.049
& 99.75
& $-0.5904$
& 0.0010 \\

& 18 & 3176 & 50000 & 45321 & 1291
& $-342.068246$
& 0.114
& 99.68
& $-0.6895$
& 0.0014 \\

& 20 & 4580 & 50000 & 46284&1860
& $-342.069334$
& 0.289
& 99.22
& $-0.6983$
& 0.0033 \\

\hline

\multirow{3}{*}{TlF}
& 12 & 619 & 20000 & 3626&200
& $-20377.137771$
& 0.003
& 99.93
& $-1.6773$
& 0.0001 \\

& 14 & 1166 & 50000 & 14861&405
& $-20377.138733$
& 0.465
& 92.60
& $-1.6813$
& 0.0018 \\

& 18 & 4551 & 50000 & 45513&1296
& $-20377.144766$
& 0.007
& 99.94
& $-1.6964$
& 0.0000 \\

& 20 & 2898 & 50000 & 48664&1991
& $-20377.144881$
& 0.008
& 99.93
& $-1.6965$
& 0.0001 \\

\end{tabular}
\end{ruledtabular}
\end{table*}
%%%%%%%%%%%%%%%%%%%%%%%%%%%%%%%%%%%%%%%%%%%%%%%%%%%%%%%%%%%%%%%%%%%%
%%%%%%%%%%%%%%%%%%%%%%%%%%%%%%%%%%%

% \subsection{Hardware results}

Beyond noiseless simulations, we benchmark the method on IBM's Marrakesh quantum processor using the AlF and TlF molecules, which uses the Heron r2 architecture, and whose native gate set consists of {CZ, RX, RZ, RZZ, SX, X}. At the time of the experiments, the device exhibited median error rates of $2.71\times10^{-3}$ (CZ), $3.53\times10^{-4}$ (SX), and $1.48\times10^{-2}$ (readout), with coherence times of $T_1=196.04~\mu$s and $T_2=98.82~\mu$s. The processor employs a heavy-hex lattice with a CLOPS (circuit layer operations per second) score of 300,000. The optimized eos-QSCI circuits are transpiled to the hardware instruction set architecture. After transpilation, the circuit depth and two-qubit gate count increase because of the limited qubit connectivity of the hardware. To reduce this overhead, we employ Qiskit's approximation degree of $0.93$, which provides an additional layer of circuit optimization before execution~\cite{javadi2024quantum}. We further apply dynamical decoupling~\cite{viola1999dynamical} and Pauli twirling~\cite{bennett1996purification} to suppress decoherence and coherent gate errors, respectively.

Table~\ref{tab:hardware_summary} summarizes the hardware results, including the number of shots and the CZ gate counts of the transpiled circuits. To improve the quality of the measured configurations, post-selection based on particle-number conservation is applied. For the 18- and 20-qubit calculations, self-consistent configuration recovery (SCCR)~\cite{robledo2025chemistry} is further employed to probabilistically reconstruct noiseless configurations. The resulting ground-state energies agree with the CASCI reference within $\sim0.01$ mHa, while the PDM deviations remain below $0.0033$ a.u. Even in 20-qubit calculations, eos-QSCI recovers $99.22\%$ and $99.93\%$ of the CASCI correlation energy for AlF and TlF, respectively, demonstrating excellent robustness against hardware noise and finite sampling.

\section{Conclusion}

In conclusion, we have developed an effective Pauli operator sorting-based QSCI framework for resource-efficient real-time evolution of the molecular electronic structure Hamiltonian. The proposed method combines fermionic Hamiltonian screening, fidelity-loss-driven weighting of Pauli operators, and cumulative weight-based Hamiltonian truncation to construct a compact effective Hamiltonian that preserves the dominant physics while significantly reducing the quantum computational cost. The resulting reduction in determinant-space dimension further lowers the cost of the classical diagonalization within the HSB-QSCI procedure. We achieve super-linear reductions in Hamiltonian terms, CX gate count, and circuit depth relative to the full Hamiltonian approach. For the 20-qubit simulations of AlF and TlF, the computed energies agree with CASCI by more than $99.99\%$, while the corresponding PDMs agree within $99.60\%$ and $99.89\%$, respectively. Notably, for TlF, these accuracies are obtained by sampling only $ 0.88\%$ of the total CASCI determinant space for the construction and diagonalization of the Hamiltonian subspace, reducing the computational overhead and capturing $90.63\%$ of the correlation energy. Furthermore, 20-qubit quantum hardware executions for both systems preserve this level of accuracy, with energy deviations below $0.01\%$ with up to $99.93\%$ of captured correlation and PDM agreements of $99.52\%$ for AlF and $99.99\%$ for TlF relative to CASCI values. We further observe that increasing the cumulative-weight threshold improves the accuracy of the computed observables.
Overall, the proposed eos-QSCI framework significantly improves the practicality of HSB-QSCI in NISQ devices by simultaneously reducing Hamiltonian complexity, quantum circuit resources, and classical post-processing costs.

\textit{Acknowledgement:} SS and AK acknowledge Dr. NM Fazil and Mr. Peniel B. Tsemo for their useful discussions on DIRAC25 package.

\textit{Data Availability Statement:} The data are available from the authors on reasonable request.

\bibliography{refs}

\clearpage
\onecolumngrid
\section*{Supplemental Material}
\appendix

\section{Reference-State-Based Screening of Hamiltonian Terms} \label{app:fermion_screening}

We analyze which second-quantized Hamiltonian terms yield non-vanishing contributions when acting on a single-determinant reference state $\ket{\Phi_0}$. This provides a structural screening criterion that eliminates operators that annihilate the reference. The electronic Hamiltonian in second quantization is given by 
\begin{equation} 
\hat{H} = \sum_{pq} h_{pq}\, a_p^\dagger a_q + \frac{1}{2} \sum_{pqrs} g_{pqrs}\, a_p^\dagger a_q^\dagger a_s a_r. 
\end{equation} 
We partition the orbital indices such that $i,j,k,l$ denote occupied orbitals in the reference determinant $\ket{\Phi_0}$, while $a,b,c,d$ denote virtual orbitals. Since the Hartree--Fock state is constructed by occupying the lowest-energy spin-orbitals, a necessary condition for a fermionic operator $\hat{O}$ to contribute is that $\hat{O}\ket{\Phi_0}\neq 0$. This requires the annihilation operators in $\hat{O}$ to act on occupied orbitals, whereas the creation operators must either act on unoccupied (virtual) orbitals or restore previously occupied orbitals.

\subsection{One-body terms}

We first consider one-body operators of the form $\hat{O}_{pq} = a_p^\dagger a_q$ and analyze their action on the reference determinant $\ket{\Phi_0}$. A non-zero contribution requires that the annihilation operator $a_q$ acts on an occupied orbital. This condition is satisfied when $q=i$. If, in addition, $p=i$, the operator $a_i^\dagger a_i$ removes and recreates an electron in the occupied space.  $a_i^\dagger a_i=n_i$, where $n_i$ is a number operator and does not generate any excited determinants. So, this term is removed.

In contrast, when $q=a$, the annihilation operator acts on an unoccupied state, and the contribution vanishes identically. The only non-trivial excitation arises for $p=a$ and $q=i$, where $a_i$ removes an electron from an occupied orbital and $a_a^\dagger$ creates one in a virtual orbital, producing a single excitation $\ket{\Phi_i^a} = a_a^\dagger a_i \ket{\Phi_0}$. All other index combinations lead to zero contributions and can therefore be discarded.

\subsection{Two-body terms}

We now consider two-body operators of the form $\hat{O}_{pqrs} = a_p^\dagger a_q^\dagger a_s a_r$. For such terms to yield non-zero contributions, both annihilation operators $a_r$ and $a_s$ must act on occupied orbitals. This condition is satisfied when $(r,s) = (i,j)$ belongs to the occupied space. If the corresponding creation operators also restore electrons within the occupied space, i.e. $(p,q) = (i,j)$, this gives the number operator and contributes to the reference energy.

More generally, when the creation operators act on virtual orbitals, $(p,q) = (a,b)$, the operator generates double excitations of the form $\ket{\Phi_{ij}^{ab}} = a_a^\dagger a_b^\dagger a_j a_i \ket{\Phi_0}$. Mixed cases, where one creation operator acts on an occupied orbital and the other on a virtual orbital, lead to single-excitation-like contributions, provided the annihilation operators act on occupied orbitals.

In contrast, any term for which at least one annihilation operator acts on a virtual orbital is zero when applied to $\ket{\Phi_0}$. This includes, for example, operators involving only virtual indices or reversed index orderings where annihilation attempts to remove electrons from unoccupied orbitals. Such terms can therefore be systematically eliminated. This screening is independent of the Hamiltonian coefficients and serves as a first-stage reduction before importance-based truncation.

% \subsection*{S1. {Cumulative operator weight and truncation error}}
%
\section{Cumulative operator weight and truncation error}  \label{app:cw_truncation}
%%%%%%%%%%%%%%%%%%%%%%%%%%%%%%%%%%%%%%%%%%%%%%%%%%%%%

\renewcommand{\thefigure}{A\arabic{figure}}
\renewcommand{\thetable}{A\arabic{table}}
\begin{figure}[H]
    \centering
    \includegraphics[width=0.95\textwidth]{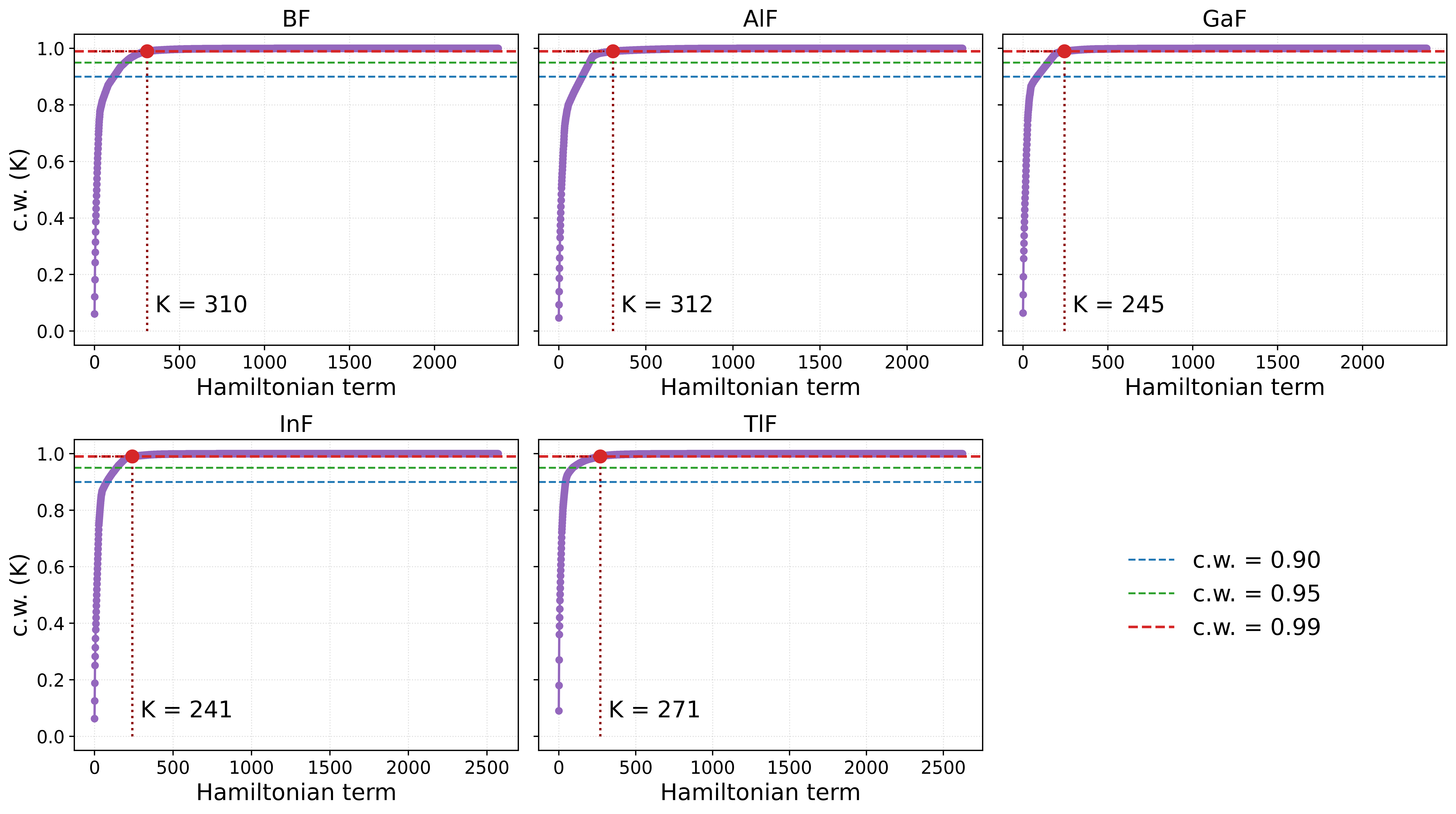}
    \caption{
Cumulative normalized weight, $c.w.(K)$, of sorted Hamiltonian term contributions for Group IIIA monofluorides for a 20-qubit system. Horizontal dashed lines indicate thresholds of $0.90$, $0.95$, and $0.99$, while the highlighted point marks the minimum number of terms $K$ required to reach $99\%$ cumulative weight to achieve chemical precision.
}
    \label{fig:20q_combine_CW}
\end{figure}

Let $\{\omega_\alpha\}_{\alpha=1}^{m}$ denote the effective Pauli operator weights, sorted in descending order,
$\omega_1 \geqslant \omega_2 \geqslant \cdots \geqslant \omega_m.$ We define the cumulative normalized operator weight as
\begin{equation}
c.w.(K) 
= \frac{W(K)}{W_{\mathrm{total}}}
= \frac{\sum_{\alpha=1}^{K} \omega_\alpha}{\sum_{\alpha=1}^{m} \omega_\alpha},
\quad K \leqslant m,
\label{eq:cw_def}
\end{equation}
where $W(K)$ and $W_{\mathrm{total}}$ denote the partial and total weights, respectively. By construction, $c.w.(K)=1$ for $K=m$. Given a threshold $\eta \in (0,1)$, we select the minimal $K$ such that $c.w. (K)= \eta,$ ensuring that a fraction $\eta$ of the total operator weight is retained. The discarded weight is defined as, $W_{\mathrm{discard}} = \sum_{\alpha=K+1}^{m} \omega_\alpha,$
which can be expressed as
\begin{align}
W_{\mathrm{discard}} 
&= W_{\mathrm{total}} - W(K) \\
&= \left[1 - c.w.(K)\right] W_{\mathrm{total}}.
\end{align}
Accordingly, the truncation error satisfies,
\begin{equation}
\delta^{(K)} 
= W_{\mathrm{discard}} 
= \left[1 - c.w.(K)\right] W_{\mathrm{total}}.
\label{eq:trunc_error}
\end{equation}

For $c.w.(K)= \eta$, the error is expressed as,
\begin{equation}
\delta^{(K)} \approx (1 - \eta)\, W_{\mathrm{total}}.
\end{equation}
In particular, for $\eta=0.99$, one obtains $\delta^{(K)} \approx 0.01\, W_{\mathrm{total}}$, indicating that only $1\%$ of the total operator weight is neglected. Thus, as $c.w. (K)$ increases, the truncation error decreases systematically.

\begin{figure}[H]
    \centering
    \includegraphics[width=1\textwidth]{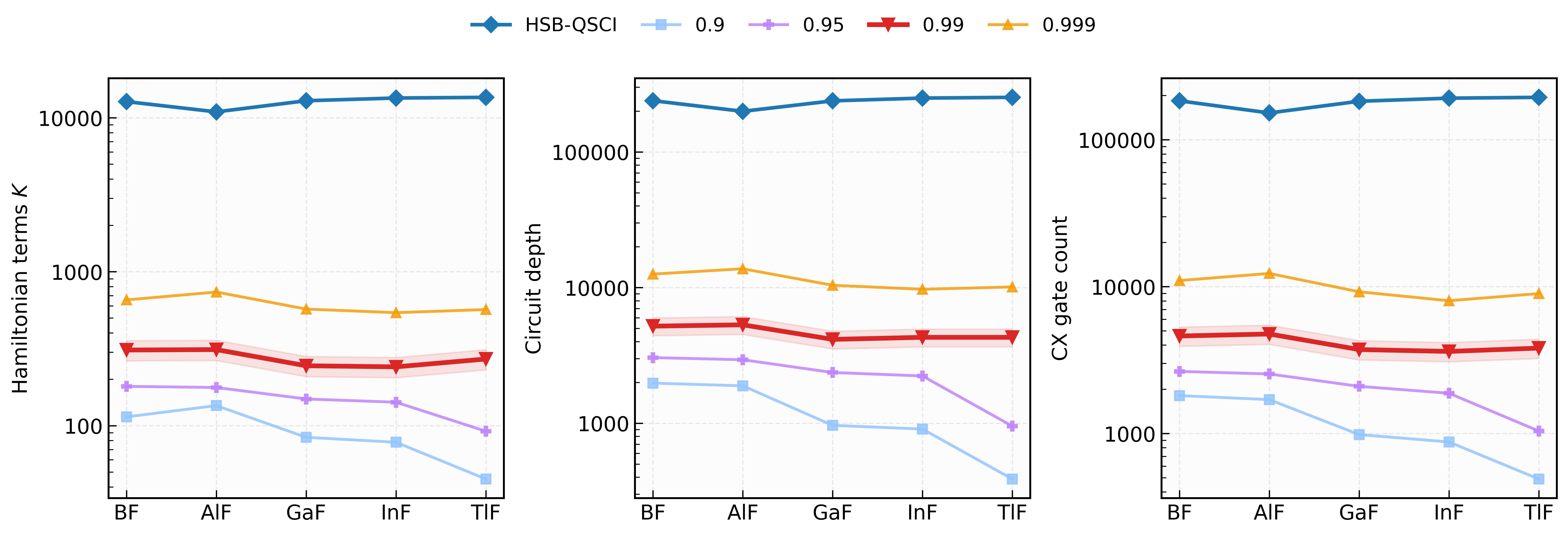}
    \caption{
Resource required for effective Hamiltonian time evolution across 20-qubit calculation of the Group IIIA monofluorides. Panels display (left) the number of retained Hamiltonian terms 
$K$, (middle) circuit depth, and (right) CX gate count at different normalized ($c.w. (K)$) threshold. The highlighted red one represents $0.99$, which is taken into consideration for all calculations.
}
    \label{fig:14q_different_CW}
\end{figure}

The numerical behavior of $c.w. (K)$ for the considered molecular systems is shown in Fig.~\ref{fig:20q_combine_CW}, where a rapid saturation of the effective Pauli operator weight demonstrates that a relatively small subset of dominant operators captures the majority of the Hamiltonian contributions. This justifies using a truncation threshold (e.g., $c.w.(K)=0.99$) in practical simulations to achieve chemical precision. The corresponding impact on resource requirements under varying thresholds is illustrated in Fig.~\ref{fig:14q_different_CW}. As the $c.w.(K)$ threshold increases, a larger number of Hamiltonian terms are retained, leading to increased quantum circuit resources. This reflects the trade-off between computational cost and accuracy, where higher thresholds result in reduced truncation error in calculating the ground-state energy as well as properties.

\section{The resource scaling of molecular systems}
\label{app:scaling}

To assess how the resource overhead of the full Hamiltonian scales relative to the eos method, we define the ratio
\begin{equation}
    R(N) = \frac{\mathcal{C}_{\mathrm{Full}}(N)}{\mathcal{C}_{\mathrm{eos}}(N)},
\end{equation}
where $\mathcal{C}$ denotes a circuit resource metric (Hamiltonian term 
count (HT), circuit depth (CD), or CX gate count (CX), and 
$N$ is the number of qubits. The eos systematically 
truncates the full Hamiltonian by sorting and retaining only the most 
significant operator contributions up to a fixed $c.w. (K)$ threshold 
(here, $99\%$).

\begin{figure}
    \centering
    \includegraphics[width=1\textwidth]{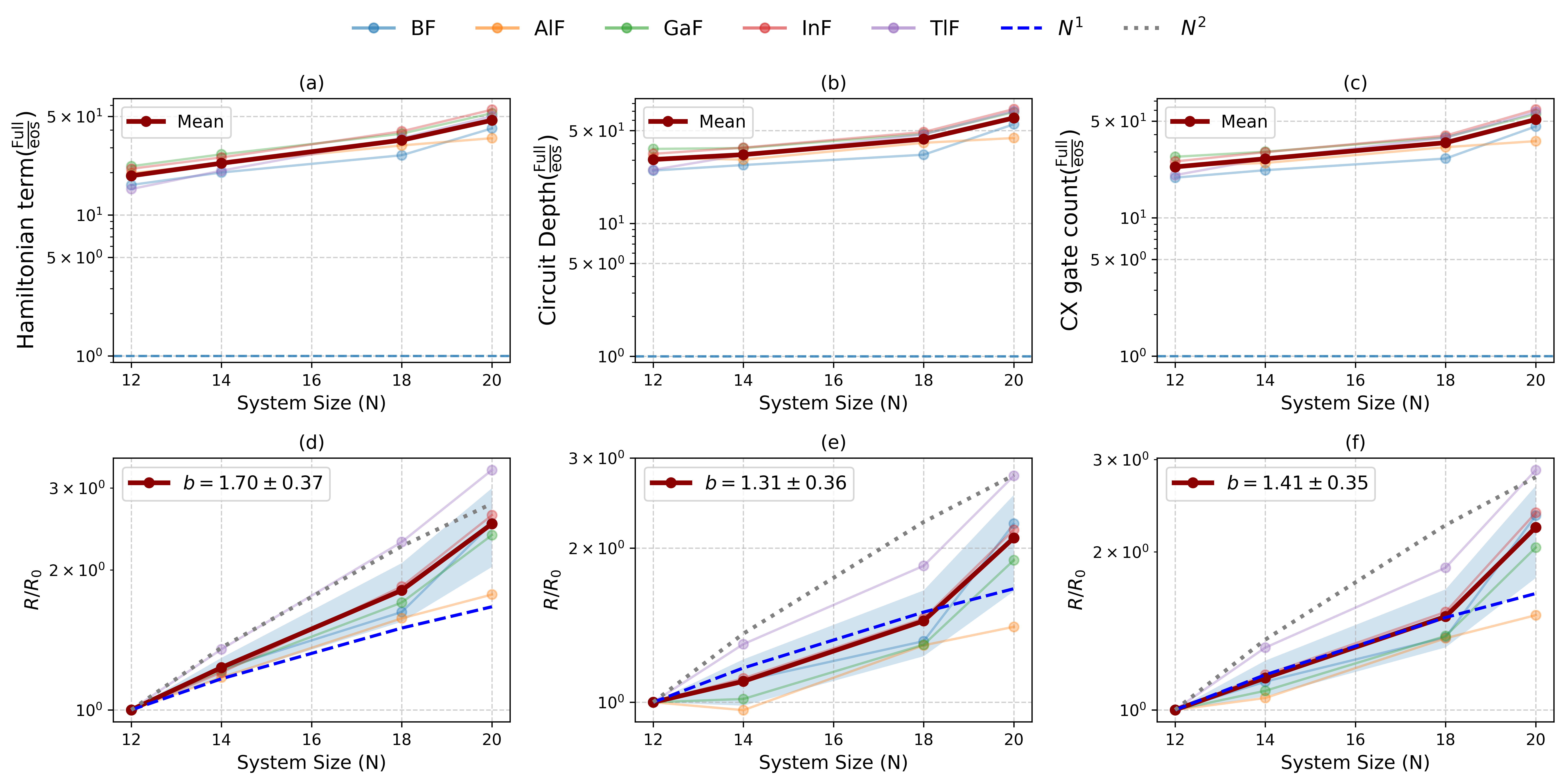}
    \caption{Finite-size scaling of resource overhead comparing the eos approach (truncated Hamiltonian) with the full Hamiltonian baseline across Group IIIA fluoride molecules. The top row (a–c) shows the raw overhead ratio $R = \mathrm{Full}/\mathrm{eos}$ for Hamiltonian term count, circuit depth, and CX gate count as a function of system size $N$. The bottom row (d–f) shows the corresponding normalized scaling collapse $R/R_0$, where $R_0$ is defined at the smallest system size. In the bottom row, the shaded region indicates one standard deviation across molecules. All panels use logarithmic scaling to highlight growth trends, and power-law fits in log–log space yield scaling exponents $b$, reported as mean $\pm$ standard deviation in the legends.}
    \label{fig:All_molecule_scaling}
\end{figure}

Fig.~\ref{fig:All_molecule_scaling} top row shows that the eos framework achieves substantial reductions in quantum computational resources compared to the full Hamiltonian approach. Across all studied molecular systems, the reduction in the number of Hamiltonian terms ranges from nearly one to two orders of magnitude, while the circuit depth and CX gate count are reduced by approximately $\sim 30\text{--}60\times$ and $\sim 20\text{--}50\times$, respectively. The horizontal reference line at $10^0 $ corresponds to the condition where the resource requirements of the full Hamiltonian and eos approaches are identical. Deviations above this line therefore quantify the degree of resource reduction achieved by the eos framework, with larger values indicating greater efficiency relative to the full Hamiltonian implementation.

To isolate the growth of this ratio within the system 
size, independent of its absolute magnitude, we normalize by the value 
at the smallest system size $N_0 = 12$:
\begin{equation}
    \widetilde{R}(N) = \frac{R(N)}{R(N_0)}.
\end{equation}

Fig.~\ref{fig:All_molecule_scaling} bottom row shows $\widetilde{R}(N)$ as a function of $N$ 
for Group-IIIA monofluorides across all 
three resource metrics. Upon normalization, the ratios exhibit a systematic and super-linear increase with system size, reaching values of approximately $\sim 2$--$4$ at $N=20$. The individual molecular trajectories are overlaid 
with their cross-molecule mean and corresponding $\pm 1\sigma$ band. To characterize the 
scaling behavior, we fit a power law $\widetilde{R}(N) \sim N^{b}$ using 
log-log linear regression, yielding
\begin{equation}
    b_{HT} = 1.70 \pm 0.37, \quad
    b_{CD} = 1.31 \pm 0.36, \quad
    b_{\mathrm{CX}} = 1.41 \pm 0.35,
\end{equation}
for the Hamiltonian term count, circuit depth, and CX gate count, respectively.

The exponents $b > 1$ obtained across all metrics indicate that the resource advantage of the eos framework grows faster than linearly with increasing system size. Physically, this behavior arises because the full Hamiltonian accumulates a rapidly increasing number of operator terms as $N$ grows, while a substantial fraction of these terms possess negligible effective operator significance, quantified through the reference-state fidelity loss criterion, and are therefore discarded within the eos screening procedure. As a result, the truncation becomes progressively more effective for larger molecular systems, yielding increasingly stronger reductions in Hamiltonian size, circuit depth, and CX gate count.

Furthermore, the observed sub-quadratic behavior ($1 < b < 2$) demonstrates that the eos framework achieves a favorable intermediate scaling regime, providing significantly enhanced resource compression. The relatively moderate standard deviations ($\Delta b \approx 0.35$) indicate some molecule-dependent variation; however, the averaged scaling exponents consistently remain within the super-linear regime across the entire monofluoride series. These results, therefore, establish that the eos implementation provides a systematically improving resource-scaling advantage over the full Hamiltonian approach as the system size increases.

%%%%%%%%%%%%%%%%%%%%%%%%%%%%%%%%%%%%%%%%%%%%%%%%%%%%%%%%%%%%%%%%%%%%%%%%%%%%%%%%%%%%%%%%%%%%

% %%%%%%%%%%%%%%%%%%%%%%%%%%%%%%%%%%%%%%%%%%%%%%%%%%%%%%%%%
\section{Ground-state energies, PDMs, and resource requirements across system sizes for Group IIIA monofluorides}\label{app:Energy_PDM_resource}

This section reports the ground-state energies, PDMs, and resource requirements for Group IIIA monofluorides across different system sizes. The eos-QSCI significantly reduces the number of Hamiltonian terms, circuit depth, and CX gates compared to HSB-QSCI. At the same time, results closely match CASCI, with only small deviations in both the ground-state energy and PDM, indicating good accuracy. Note that all the calculations are carried out at $c.w. (K)=0.99$.
%%%%%%%%%%%%%%%%%%%%%%%%%%%%%%%%%%%%%%%%%%%%%%%%%%%%%%%%%%%%
%BF table%
% \FloatBarrier

\begin{table*}
\setlength{\tabcolsep}{6pt}
\centering
\caption{Ground state energies and PDMs of BF molecule for different system sizes. Energies are in Hartree (Ha) and PDM in atomic units (a.u.). Absolute deviation is defined as $\Delta X = |X_{\mathrm{eos\text{-}QSCI}} - X_{\mathrm{CASCI}}|$, where $X \in \{E, PDM\}$ represents the difference between the exact CASCI diagonalization and the eos-QSCI approach.}
\begin{tabular}{rrcccc}
\toprule
\toprule
System size & Method 
& Energy & PDM & $\Delta E$ & $\Delta PDM$ \\
\midrule

\multicolumn{2}{r}{DHF} 
& -124.267891 & 0.3416 & -- & -- \\
\midrule

\multirow{2}{*}{12}
& CASCI 
& \multicolumn{2}{c}{-124.272369 \quad 0.3358} & \multicolumn{2}{c}{} \\
& eos-QSCI 
& -124.272125 & 0.3371 & 0.000244 & 0.0013 \\
\midrule

\multirow{2}{*}{14}
& CASCI 
& \multicolumn{2}{c}{-124.285105 \quad 0.3015} & \multicolumn{2}{c}{} \\
& eos-QSCI 
& -124.284758 & 0.3005 & 0.000347 & 0.0010 \\
\midrule

\multirow{2}{*}{18}
& CASCI 
& \multicolumn{2}{c}{-124.304952 \quad 0.2186} & \multicolumn{2}{c}{} \\
& eos-QSCI 
& -124.304067 & 0.2109 & 0.000885 & 0.0077 \\
\midrule

\multirow{2}{*}{20}
& CASCI 
& \multicolumn{2}{c}{-124.305001 \quad 0.2193} & \multicolumn{2}{c}{} \\
& eos-QSCI 
& -124.304144 & 0.2144 & 0.000857 & 0.0049 \\
\bottomrule
\bottomrule
\end{tabular}
\label{tab:BF_energy-pdm}
\end{table*}

%AlF table%
\begin{table*}
\setlength{\tabcolsep}{6pt}
\centering
\caption{Ground-state energies and PDMs of AlF molecule for different system sizes.}
\begin{tabular}{rrcccc}
\toprule
\toprule
System size & Method
& Energy & PDM & $\Delta E$ & $\Delta$PDM \\
\midrule

\multicolumn{2}{r}{DHF}
& -342.032557 & -0.5283 & -- & -- \\
\midrule

\multirow{2}{*}{12}
& CASCI
& \multicolumn{2}{c}{-342.041647 \quad -0.5565} & \multicolumn{2}{c}{} \\
& eos-QSCI
& -342.041328 & -0.5480 & 0.000319 & 0.0085 \\
\midrule

\multirow{2}{*}{14}
& CASCI
& \multicolumn{2}{c}{-342.053050 \quad -0.5894} & \multicolumn{2}{c}{} \\
& eos-QSCI
& -342.052734 & -0.5745 & 0.000316 & 0.0149 \\
\midrule

\multirow{2}{*}{18}
& CASCI
& \multicolumn{2}{c}{-342.068360 \quad -0.6881} & \multicolumn{2}{c}{} \\
& eos-QSCI
& -342.067476 & -0.6796 & 0.000884 & 0.0085 \\
\midrule

\multirow{2}{*}{20}
& CASCI
& \multicolumn{2}{c}{-342.069623 \quad -0.6950} & \multicolumn{2}{c}{} \\
& eos-QSCI
& -342.068686 & -0.6976 & 0.000937 & 0.0026 \\
\bottomrule
\bottomrule
\end{tabular}
\label{tab:AlF_energy-pdm}
\end{table*}

%GaF table%
\begin{table*}
\setlength{\tabcolsep}{6pt}
\centering
\caption{Ground state energies and PDMs of GaF molecule for different system sizes.}
\begin{tabular}{rrcccc}
\toprule
\toprule
System size & Method 
& Energy & PDM & $\Delta E$ & $\Delta PDM$ \\
\midrule

\multicolumn{2}{r}{DHF} 
& -2042.239419 & -0.9199 & -- & -- \\
\midrule

\multirow{2}{*}{12}
& CASCI 
& \multicolumn{2}{c}{-2042.245573 \quad -0.9126} & \multicolumn{2}{c}{} \\
& eos-QSCI 
& -2042.245340 & -0.9115 & 0.000233 & 0.0011 \\
\midrule

\multirow{2}{*}{14}
& CASCI 
& \multicolumn{2}{c}{-2042.253725 \quad -0.8979} & \multicolumn{2}{c}{} \\
& eos-QSCI 
& -2042.253392 & -0.8951 & 0.000333 & 0.0028 \\
\midrule

\multirow{2}{*}{18}
& CASCI 
& \multicolumn{2}{c}{-2042.265486 \quad -0.8455} & \multicolumn{2}{c}{} \\
& eos-QSCI 
& -2042.264692 & -0.8432 & 0.000794 & 0.0023 \\
\midrule

\multirow{2}{*}{20}
& CASCI 
& \multicolumn{2}{c}{-2042.265922 \quad -0.8432} & \multicolumn{2}{c}{} \\
& eos-QSCI 
& -2042.264856 & -0.8449 & 0.001066 & 0.0017 \\
\bottomrule
\bottomrule
\end{tabular}
\label{tab:GaF_energy-pdm}
\end{table*}

%InF table%

\begin{table*}[t]
\setlength{\tabcolsep}{6pt}
\centering
\caption{Ground state energies and PDMs of InF molecule for different system sizes.}
\begin{tabular}{rrcccc}
\toprule
\toprule
System size & Method 
& Energy & PDM & $\Delta E$ & $\Delta PDM$ \\
\midrule

\multicolumn{2}{r}{DHF} 
& -5980.216422 & -1.2564 & -- & -- \\
\midrule

\multirow{2}{*}{12}
& CASCI 
& \multicolumn{2}{c}{-5980.224186 \quad -1.2423} & \multicolumn{2}{c}{} \\
& eos-QSCI 
& -5980.223868 & -1.2430 & 0.000318 & 0.0007 \\
\midrule

\multirow{2}{*}{14}
& CASCI 
& \multicolumn{2}{c}{-5980.231301 \quad -1.2232} & \multicolumn{2}{c}{} \\
& eos-QSCI 
& -5980.230966 & -1.2213 & 0.000335 & 0.0019 \\
\midrule

\multirow{2}{*}{18}
& CASCI 
& \multicolumn{2}{c}{-5980.241354 \quad -1.1632} & \multicolumn{2}{c}{} \\
& eos-QSCI 
& -5980.240523 & -1.1626 & 0.000831 & 0.0006 \\
\midrule

\multirow{2}{*}{20}
& CASCI 
& \multicolumn{2}{c}{-5980.241773 \quad -1.1604} & \multicolumn{2}{c}{} \\
& eos-QSCI 
& -5980.240780 & -1.1581 & 0.000993 & 0.0023 \\
\bottomrule
\bottomrule
\end{tabular}
\label{tab:InF_energy-pdm}
\end{table*}

%%% TlF table %%%%%%%%%%

\begin{table*}
\setlength{\tabcolsep}{6pt}
\centering
\caption{Ground-state energies and PDMs of TlF molecule for different system sizes.}
\begin{tabular}{rrcccc}
\toprule
\toprule
System size & Method
& Energy & PDM & $\Delta E$ & $\Delta$PDM \\
\midrule

\multicolumn{2}{r}{DHF}
& -20377.132915 & -1.6725 & -- & -- \\
\midrule

\multirow{2}{*}{12}
& CASCI
& \multicolumn{2}{c}{-20377.137774 \quad -1.6774} & \multicolumn{2}{c}{} \\
& eos-QSCI
& -20377.137310 & -1.6750 & 0.000464 & 0.0024 \\
\midrule

\multirow{2}{*}{14}
& CASCI
& \multicolumn{2}{c}{-20377.139198 \quad -1.6831} & \multicolumn{2}{c}{} \\
& eos-QSCI
& -20377.138696 & -1.6776 & 0.000502 & 0.0055 \\
\midrule

\multirow{2}{*}{18}
& CASCI
& \multicolumn{2}{c}{-20377.144773 \quad -1.6964} & \multicolumn{2}{c}{} \\
& eos-QSCI
& -20377.143590 & -1.6923 & 0.001183 & 0.0041 \\
\midrule

\multirow{2}{*}{20}
& CASCI
& \multicolumn{2}{c}{-20377.144889 \quad -1.6966} & \multicolumn{2}{c}{} \\
& eos-QSCI
& -20377.143768 & -1.6896 & 0.001121 & 0.0070 \\
\bottomrule
\bottomrule
\end{tabular}
\label{tab:TlF_energy-pdm}
\end{table*}
%%%%%%%%%%%%%%%%%%%%%%%%%%%%%%%%%%%%%%%%%%%%%%%%%%%%%%%%%%%%

\begin{table*}[t]
\centering
\small
\setlength{\tabcolsep}{5pt}
\renewcommand{\arraystretch}{1.1}

\caption{
Quantum resource requirements for the BF molecule using Full and eos methods.
HT denotes Hamiltonian terms, CD is circuit depth, and CX represents the number of CX gates.
The percentage reduction is computed relative to the full Hamiltonian.
}

\label{tab:BF-resources}

\begin{ruledtabular}
\begin{tabular}{c|ccc|ccc|ccc}
\multirow{2}{*}{System Size}
& \multicolumn{3}{c|}{HT}
& \multicolumn{3}{c|}{CD}
& \multicolumn{3}{c}{CX} \\
& Full
& eos
& \% Red.
& Full
& eos 
& \% Red.
& Full
& eos 
& \% Red. \\
\hline

12
& 1535
& 94
& 93.87
& 20571
& 822
& 96.00
& 14020
& 718
& 94.87 \\

14
& 2982
& 149
& 95.00
& 44248
& 1606
& 96.37
& 31502
& 1425
& 95.47 \\

18
& 8088
& 305
& 96.22
& 140419
& 4263
& 96.96
& 105946
& 3947
& 96.27 \\

20
& 12747
& 310
& 97.56
& 238826
& 4274
& 98.21
& 184132
& 4019
& 97.81 \\

\end{tabular}
\end{ruledtabular}
\end{table*}

%%%%%%%%%%%%%%%%%%%%%%%%%%%%%%%%%%%%%%%%%%%%%%%%%%%%%%%%%%%

%%%%%%%%%%%%%%%%%%%%%%%%%%%%%%%%%%%%%%%%%%%%%%%%%%%%%%%%%%%%%%%%%%%%%%%

\begin{table*}[t]
\centering
\small
\setlength{\tabcolsep}{5pt}
\renewcommand{\arraystretch}{1.1}

\caption{
Quantum resource requirements for the GaF molecule using the eos method.
The eos method values correspond to the optimized circuits.
The percentage reduction is computed relative to the full Hamiltonian.
}

\label{tab:GaF-resources}

\begin{ruledtabular}
\begin{tabular}{c|ccc|ccc|ccc}
\multirow{2}{*}{System Size}
& \multicolumn{3}{c|}{HT}
& \multicolumn{3}{c|}{CD}
& \multicolumn{3}{c}{CX} \\
& Full
& eos 
& \% Red.
& Full
& eos 
& \% Red.
& Full
& eos 
& \% Red. \\
\hline

12
& 1663
& 75
& 95.49
& 22508
& 618
& 97.25
& 15396
& 557
& 96.38 \\

14
& 3310
& 123
& 96.28
& 49744
& 1346
& 97.29
& 35470
& 1180
& 96.67 \\

18
& 8904
& 236
& 97.34
& 154997
& 3285
& 97.88
& 116834
& 3056
& 97.38 \\

20
& 12913
& 245
& 98.10
& 238110
& 3450
& 98.55
& 182896
& 3245
& 98.22 \\

\end{tabular}
\end{ruledtabular}
\end{table*}

%%%%%%%%%%%%%%%%%%%%%%%%%%%%%%%%%%%%%%%%%%%%%%%%%%%%%%%%%%%%%%%%%%%%%%%%%%%%%%%%%%%%%%%%

\begin{table*}[t]
\centering
\small
\setlength{\tabcolsep}{5pt}
\renewcommand{\arraystretch}{1.1}

\caption{
Quantum resource requirements for the InF molecule using the eos method.
The eos values correspond to the optimized circuits.
The percentage reduction is computed relative to the full Hamiltonian.
}

\label{tab:InF-resources}

\begin{ruledtabular}
\begin{tabular}{c|ccc|ccc|ccc}
\multirow{2}{*}{System Size}
& \multicolumn{3}{c|}{HT}
& \multicolumn{3}{c|}{CD}
& \multicolumn{3}{c}{CX} \\
& Full
& eos 
& \% Red.
& Full
& eos 
& \% Red.
& Full
& eos 
& \% Red. \\
\hline

12
& 1723
& 81
& 95.29
& 23385
& 696
& 97.02
& 15980
& 625
& 96.08 \\

14
& 3378
& 132
& 96.09
& 50728
& 1362
& 97.31
& 36102
& 1210
& 96.64 \\

18
& 9210
& 235
& 97.44
& 161243
& 3306
& 97.94
& 121654
& 3096
& 97.45 \\

20
& 13447
& 241
& 98.20
& 249556
& 3433
& 98.62
& 191956
& 3159
& 98.19 \\

\end{tabular}
\end{ruledtabular}
\end{table*}

\end{document}